\documentclass[aps,prl,twocolumn,superscriptaddress,showpacs]{revtex4}

\usepackage{graphicx}

\begin{document}

\title{Collective effects in vortex movements in complex (dusty) plasmas}
\author{Mierk Schwabe}
\email[]{mierk@berkeley.edu}
%\thanks{}
\affiliation{Department of Chemical and Biomolecular Engineering,
  University of California, Berkeley, CA 94720, USA}
\affiliation{Max Planck Institute for extraterrestrial Physics, PO Box 1312, Giessenbachstr.,
   85741 Garching, Germany}
\author{Sergey Zhdanov}
\affiliation{Max Planck Institute for extraterrestrial Physics, PO Box 1312, Giessenbachstr.,
   85741 Garching, Germany}
 \author{Christoph R\"{a}th}
 \affiliation{Max Planck Institute for extraterrestrial Physics, PO Box 1312, Giessenbachstr.,
    85741 Garching, Germany}
\author{David~B.~Graves}
\affiliation{Department of Chemical and Biomolecular Engineering,
  University of California, Berkeley, CA 94720, USA}
\author{Hubertus M. Thomas}
\affiliation{Max Planck Institute for extraterrestrial Physics, PO Box 1312, Giessenbachstr.,
   85741 Garching, Germany}
\author{Gregor E. Morfill}
\affiliation{Max Planck Institute for extraterrestrial Physics, PO Box 1312, Giessenbachstr.,
   85741 Garching, Germany}

\date{\today}

\begin{abstract}
We study the onset and characteristics of vortices in complex (dusty)
plasmas using two-dimensional simulations in a setup modeled after the
PK-3 Plus laboratory. A small number of microparticles initially
self-arranges in a monolayer around the void. As additional particles
are introduced, an extended system of vortices develops due to a non-zero curl of the plasma forces. We demonstrate a
shear-thinning effect in the vortices. Velocity structure functions and the
energy and enstrophy spectra show that vortex flow turbulence is
present that is in essence of the ``classical'' Kolmogorov type.
\end{abstract}

%\keywords{}

\maketitle

{\it Introduction---}Complex plasmas consist of micro\-meter-sized particles embedded in a
low temperature plasma. Experimenters can record the trajectories of
the microparticles with digital cameras and study the microparticle
behavior in detail. Complex plasmas are ideal model
systems for nano fluids, phase transitions, transport processes, etc \cite{Morfill2009}. They display many
collective effects. For instance, vortices in which the
microparticles move in circles are common. They appear when gravity is
compensated by thermophoresis
\cite{Morfill2004b,Zuzic2007,Schwabe2009} and in weightless systems
\cite{Morfill1999a,Fortov2003,Nefedov2003} and can be induced
externally, for instance by a laser \cite{Klindworth2000,Miksch2007},
by a probe \cite{Law1998,Uchida2009}, or by gas flow
\cite{Vladimirov2001,Mitic2008,Schwabe2011b,Fink2013}. Vortices can convoy
nucleation \cite{Nosenko2007} and lead to shear flow instabilities at
their edges \cite{Heidemann2011a}. The origin of self-excited vortices
in complex plasmas has long been a topic of discussion. They could be
induced by the microparticle size dispersion
\cite{Zuzic2007,Yokota1996}, charge gradients
\cite{Vaulina2000,Vaulina2003}, the void \cite{Mamun2004}, or a non-zero curl of the forces that the plasma
exerts on the microparticles \cite{Akdim2003,Goedheer2003}.

As we show in the following, vortices in complex plasmas are an ideal test-bed for studying the onset of turbulence and collective effects
on the microcanonical level \cite{Morfill2004b}. The
description and dynamics of a turbulent complex plasma is an outstanding
problem. Turbulence can be present in the background gas
\cite{Robertson2007,Schwabe2011b} or plasma
\cite{Klinger2001,Benkadda2002} or induced by instabilities or waves
in the dusty plasma itself \cite{Shukla2004,Pramanik2003,Tsai2012}. In comparison with traditional experiments on turbulence
(see, e.g., \cite{Lewis1999,Arneodo2008,Monchaux2012}), the particles that transmit the
interaction can be visualized directly. Still many physical questions
remain unanswered partly because the adequate simulations are
lacking. Simulations help identify relevant experimental parameters
and lead the way to designing related experiments.

Hybrid simulations are especially well suited to study self-excited
vortices. In these simulations, the plasma is typically simulated as a
fluid and coupled with a separate simulation of the microparticle
dynamics. This allows taking into account the forces that the plasma
exerts on the microparticles and letting vortices develop
naturally. Then, the reason for the self-excitation of vortices and
the exact process of the onset of vortices can be
determined.

Here, we present two-dimensional simulations of vortices in complex
plasmas to investigate the onset of turbulence and collective
effects. We model the discharge plasma and the cooperative dynamics of
the microparticles embedded in it and demonstrate that the
microparticle shear flow is relevant to that observed experimentally,
and subjected to shear-thinning and turbulence.

{\it Simulation particulars---}We use the simulation setup described in
\cite{Schwabe2013a}, which is a two-dimensional model of the central
plane of the PK-3 Plus chamber \cite{Thomas2008}. It consists of a
hybrid fluid-analytical description of the plasma by \citet{Kawamura2011}, coupled to a molecular-dynamics (MD) simulation of the microparticle
dynamics. The MD simulation is based on the Large-scale
Atomic/Molecular Massively Parallel Simulator (LAMMPS) \cite{Plimpton1995},
modified to take into account the plasma forces. The microparticles
interact via Yukawa pair potentials and are
subject to random forces mimicking interactions with background atoms bumping into
them. There is also a background friction force exerted by the neutral gas on the
microparticles, the so-called Epstein drag force \cite{Epstein1924}. The ion flux and ambipolar electric field from the
plasma simulation are used to calculate the ion drag and
confinement forces acting on the microparticles. A naturally occurring
non-vanishing curl of the sum of these forces leads to the generation of vortices.

For the simulations presented here, we use argon at a pressure of 10\,Pa as
buffer gas. The microparticles have a diameter of 3.4\,$\mu$m, a mass
of $3.1 \times 10^{-14}$\,kg, and a
charge of -3481\,e. The friction constant with the background gas is
$\gamma_{\text{Ep}} = 37\,\text{s}^{-1}$. The electron temperature is 2.4\,eV, the ion and
neutral gas temperatures 300\,K. The mean electron and ion number densities
are $3.4 \times 10^{14}\,\text{m}^{-3}$, with a maximum in the center of
the simulation box (see \cite{Schwabe2013a}). We do not take into account the
effect of the microparticles on the plasma. Unless otherwise
mentioned, we used 7100 particles in the simulations runs presented here. The areal density of the cloud is
$1.4 \times 10^7\,\text{m}^{-2}$.

\begin{figure}[]
  \includegraphics[width=.9\columnwidth]{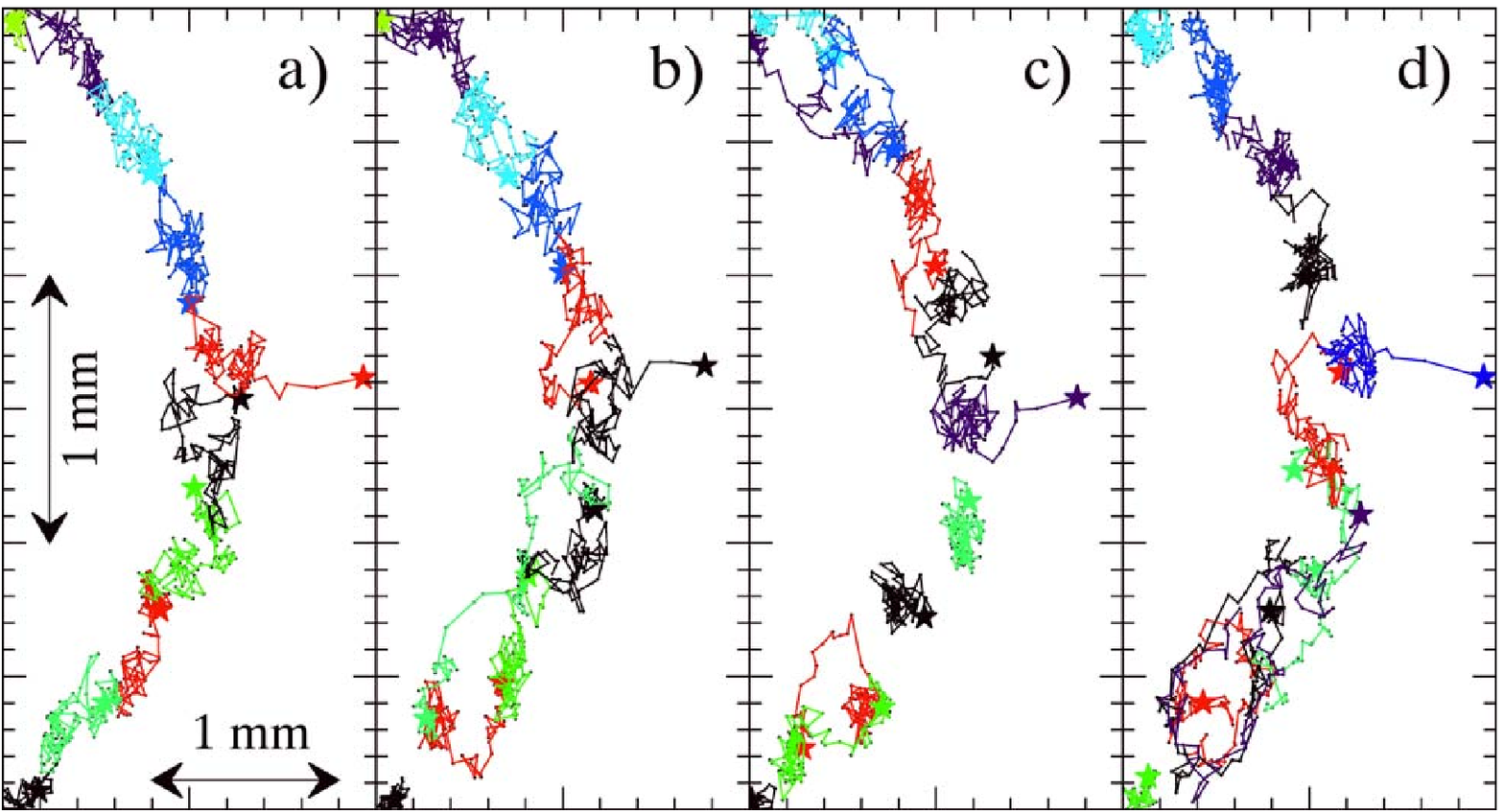}\\
  \includegraphics[width=.9\columnwidth]{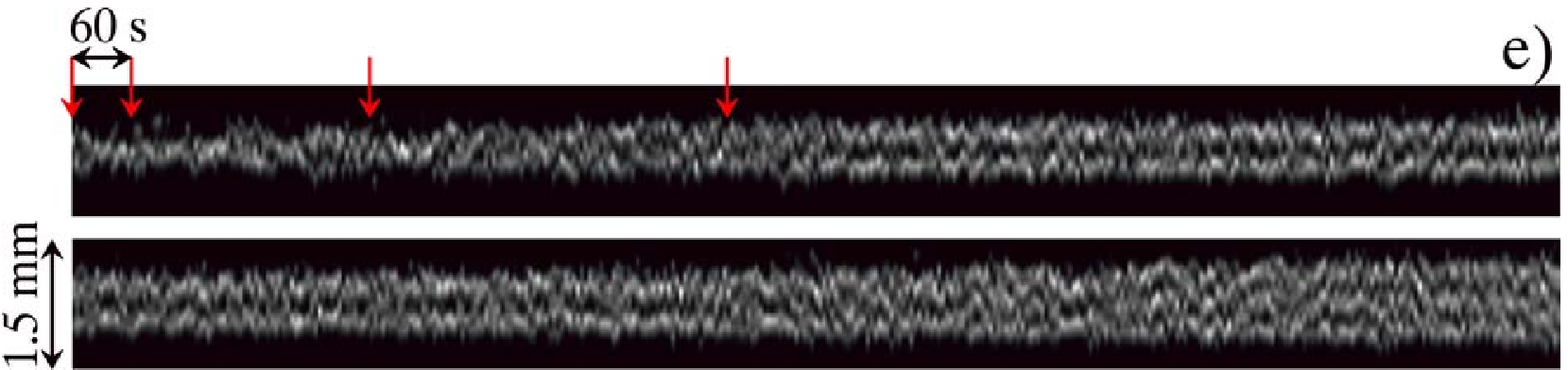}
  \caption{\label{fig:traj} (color online) (a-d) Trajectories that show the process of
    adding single particles to a monolayer. The trajectories of
    individual particles are distinguished by color. The start point
    of each trajectory is indicated by a star. e) Space-time plot (periodgram) that demonstrates
    the development of the width of the central part of the
    microparticle cloud with time. To make this picture, we added the pixel intensities of
    images showing the particle positions in the region $y = 9.5 -
    10.5$\,mm as a function of $x$ position for frames corresponding to
    a total of 50 particle injections and plot the resulting graphs
    next to each other. In the periodgram, the vertical
    axis corresponds to $x$-position, and the horizontal axis to
    time. Injections occur every minute. The arrows indicate the injections a)-d).}
\end{figure}

{\it Onset of vortices---}We mimic the experimental injection process by placing microparticles
near the sheath edge. They are then pushed into the plasma by the
ambipolar electric field. Inside the plasma bulk, the ion drag force
drives the microparticles around the void, where they arrange in a
monolayer along the equilibrium line where the counteracting ion drag
and ambipolar electric force are balanced. Figure~\ref{fig:traj} shows trajectories of particles inside this monolayer
as progressively more particles are added. %(see \cite{movie} for a movie).
A single particle added to
the monolayer rapidly finds a place inside
(Fig.~\ref{fig:traj}(a)). The next added particle is also incorporated
into the monolayer, but the propagation of forces
    along the monolayer leads to a particle further down (shown in
    light green) being ejected from the layer. The ejected particle
    hops along the inside of the monolayer
    (Fig.~\ref{fig:traj}(b)). The next particles that are
    injected disturb the layer, causing transverse displacement of other
    particles, but no vortices appear yet (Fig.~\ref{fig:traj}(c)). Two layers of particles form, and vortex
    motion ensues only after the 12th particle injection
    (Fig.~\ref{fig:traj}(d)). The space-time plot (periodgram) (Fig.~\ref{fig:traj}(e)) demonstrates the growth of the cloud width with additional injections of single particles. An additional particle is injected every 60\,s.

\begin{figure}
  \includegraphics[width=.9\columnwidth]{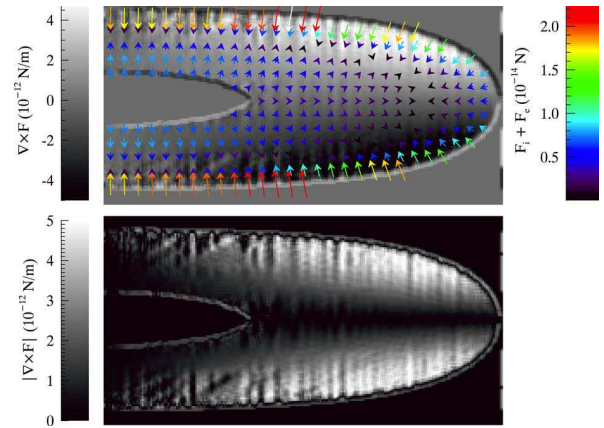}
  \caption{\label{fig:Fie}(color online) (top) The sum of the ion drag $F_i$ and electric
    force $F_e$ acting on the particles (vectors, color online, right
    color map) overlaid a map of the curl of the force field (greyscale, left color map). The equilibrium line is
    clearly visible in the vector field. The edge of the cloud in the
    curl map is enhanced, which is an artefact of the way curl F was
  calculated. (bottom) Map of the absolute value of the curl of the
  force field. Both curl maps uses a threshold of $\pm 5 \times 10^{-12}$\,N/m to enhance the
contrast.}
\end{figure}

Figure~\ref{fig:Fie} shows the vector plot of the combined plasma
forces acting on the particles, the ion drag force $F_i$ and the
ambipolar electric force $F_e$. There is a single line around the void at which
the sum of these two forces vanishes. This equilibrium line is visible
in Fig.~\ref{fig:Fie} as the line to which the vectors point. On the
outside of the equilibrium line, the confining ambipolar electric force
dominates, whereas on the inside of the equilibrium line, the ion drag force
dominates, keeping the central 'void' particle-free. The cause of the
finite width of the microparticle cloud is the mutual interparticle
repulsion.

The gray-scaled background of Fig.~\ref{fig:Fie} maps the non-zero component of
the curl of the plasma force field, $\nabla \times ({\bf F_i} + {\bf
  F_e})$ \cite{footnote1}. The magnitude of the curl of the vector field is largest near the
edges of the microparticle cloud, on the outside of the equilibrium
line, whereas it nearly vanishes on the inside of the equilibrium line and in
the mid-plane of the simulation box. This qualitatively confirms the basic result of
\citet{Akdim2003}: The vortices can be caused by a non-vanishing curl
of the plasma force (see also \cite{footnote2}).

\begin{figure}
  \includegraphics[width=.9\columnwidth]{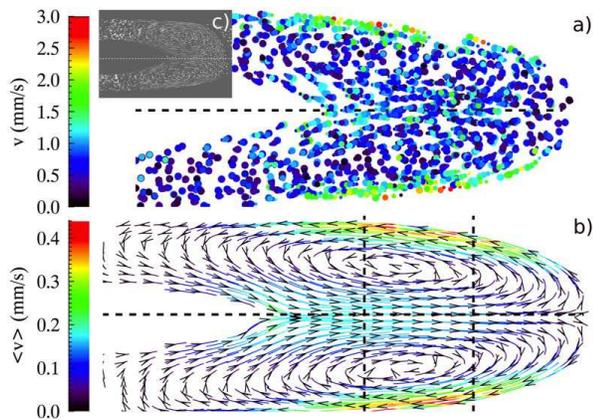}
  \caption{\label{fig:velocities}(color online) Simulated (a,b) and
    experimental vortices (c). The horizontal dashed lines mark the chamber
    midplane/simulation box center. The field of view is (a,b) $23
    \times 10$\,mm$^2$ and (c)
    $30 \times 20$\,mm$^2$. a)  Positions of 1/10 of all particles
    in the simulation. The dots indicate four particle positions 0.2\,s apart
    for each particle and grow with time. The color of the dots indicates
    the total particle velocity, which is shown with a threshold of
    3\,mm/s. b) Streamlines and arrows indicate the mean
    velocity field of the simulated vortices. The mean was calculated
    by averaging over at least five particles for each position. The
    vertical dashed lines indicate the position of the sliding window
    used for Fig.~\ref{fig:ratestress}.
    c) Overlay of 60 pictures (30\,s) taken with the PK-3 Plus
    laboratory, courtesy of the PK-3 Plus team \cite{Thomas2008}. The images are mirrored horizontally to correspond to the simulation.}
\end{figure}

{\it Fully developed system of vortices---}Once several layers of microparticles are present, the vortices are
fully developed. Figure~\ref{fig:velocities} visualizes the movement of
microparticles inside a cloud with two vortices. The particles move towards the
simulation box center along the outside of the cloud and back towards the
cloud edge in the central plane, as in the experiments in the PK-3
Plus laboratory on board the International Space Station \cite{footnote3}.

\begin{figure}
  \includegraphics[width=.9\columnwidth]{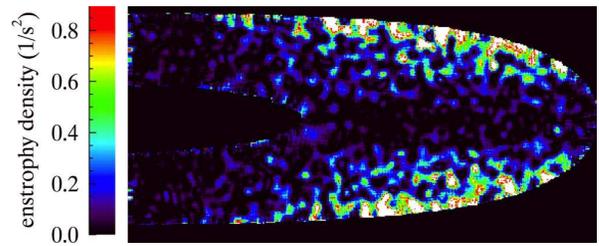}
  \caption{\label{fig:enstrophy}(color online) Enstrophy density
    $\Omega$ of the vortices. Field of view: $23 \times 12\,\text{mm}^2$.}
\end{figure}

The particles move fastest at the edge of
the vortices, where the rotational energy is concentrated, see
Fig.~\ref{fig:enstrophy} showing the enstrophy density of system. We calculate the enstrophy density
following \cite{Heidemann2011a}: Firstly, we determine the 16 closest
neighbors of every particle $i$ and find the vectors to their
positions $\bf{r_{ij}} = \bf{r_j} - \bf{r_i}$ and the relative
velocities $\bf{v_{ij}} = \bf{v_j} - \bf{v_i}$. The vorticity
  $\omega_i$ is a
  function of the projections $c_{ij}$ of the relative velocity onto the
  vectors orthogonal to $\bf{r_{ij}}$:
 \begin{equation}
   \omega_i = {\bf e_z} \cdot (\text{\bf curl}\, {\bf v})_i = \frac{1}{nn} \sum_{j=1}^{nn}c_{ij},
\end{equation}
 where $nn = 16$ is the number of neighbors used in the calculation,
 $\bf{e_z} = \bf{e_x} \times \bf{e_y}$ is the unit vector in direction
 perpendicular to the plane of the simulation, and $c_{ij} = ({\bf e_z}
 \times {\bf r_{ij}}) \cdot {\bf v_{ij}} / r_{ij}^2$. Finally, we calculate the enstrophy density $\Omega_{kl}$ in every grid cell (k,l)
from the average squared vorticity
\begin{equation}
  \Omega_{kl} = \left(\frac{1}{n} \sum_{i=0}^n \omega_i\right)^2,
\end{equation}
where n is the number of particles in the cell.

Figure~\ref{fig:enstrophy} shows that the enstrophy density is
concentrated at the edges of the microparticle cloud, where the
microparticles move the fastest. As there are considerable velocity
differences across the cloud, we shall next see whether the vortices
produce any shear stress.

\begin{figure}[b]
  \includegraphics[width=.9\columnwidth]{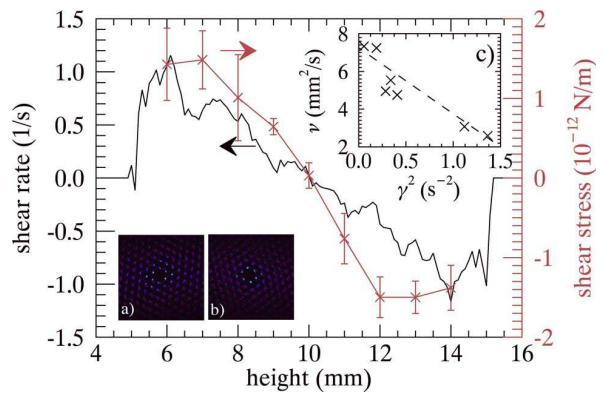}
  \caption{\label{fig:ratestress}(color online) Shear rate and shear stress as a
    function of height above the lower sheath edge. Insets: Radial
    pair correlation functions at the lower (a) and upper (b) edges of
    the cloud, c) viscosity $\nu$ as a function of squared shear stress
    $\gamma^2$, demonstrating shear thinning. The dashed line is a
    guide to the eye.}
\end{figure}

{\it Shear thinning---}We follow \cite{Nosenko2013} to calculate the shear stress $\sigma_{xy}$, i.e.,
the off-diagonal elements of the pressure tensor, that is build up from
the contributions of each particle's neighbors, neglecting the kinetic
component. For a two-dimensional system in polar coordinates,
$\sigma_{xy}$ is given by
\begin{equation}
  \sigma_{xy} = - \frac{1}{2}n^2 \int_0^{r_{max}} dr \, r^2 \frac{dV}{dr}
  \int_0^{2 \pi} d \phi \cos \phi \sin \phi \, g(r,\phi),
\end{equation}
where $n$ is the number density of the microparticles, $V$ is the
Yukawa potential, and $g(r,\phi)$ is the radial pair distribution
function. The insets (a) and (b) in Fig.~\ref{fig:ratestress} show the
pair correlation functions at the upper and lower edges of the
cloud. As in \cite{Nosenko2013}, the elliptical distortion of
$g(r,\phi)$ shows strong shear stress.

The main part of Fig.~\ref{fig:ratestress} shows the shear rate $\gamma = (\partial
v_x /  \partial y) + (\partial v_y / \partial x)$, calculated from
the velocity maps, and the shear stress $\sigma_{xy}$ as a function of
height above the lower sheath edge in a sliding window of size $5
\times 2\,\text{mm}^2$ at a horizontal distance of 12\,mm from the
plasma center (see Fig.~\ref{fig:velocities}). The vertical center of
the simulation box is at a height of 10\,mm. Both the shear rate and the shear stress vanish in the
middle, where all particles are moving towards the edge of the
cloud. Their absolute values are largest close to the area where the
opposite fluxes induced by the vortices meet, as
expected. Numerically, the values are close to those found in
experiments by \citet{Nosenko2013}, in which laser-induced motion of
microparticles is studied. We also observe shear-thinning, as can be seen in
Fig.~\ref{fig:ratestress}(c) \cite{footnote4}: The viscosity $\nu = \sigma_{xy}/(\rho
\gamma)$, with $\rho$ being the areal mass density, decreases as a
function of squared shear stress.

{\it Flow velocity---}The parameters of the simulated vortex flow fit real experiments performed with
complex plasmas. For instance, using a mean viscosity of $\nu = 5\,\text{mm}^2\text{/s}$, a flow velocity of
$v = 2\,\text{mm/s}$ and a typical size of the vortices of $L =
10\,\text{mm}$ (see Figs.~\ref{fig:velocities},\ref{fig:ratestress}), we obtain a
Reynolds number $\mathcal{R} = v L / \nu \sim 4$, which is well
in-line with observations \cite{Morfill2004b,Heidemann2011a}. At such
a low value, e.g., water flow is typically laminar. Nevertheless, we do find evidence that turbulence
is present in the system of vortices, as we shall see next.

{\it Turbulence---}The origin of the turbulent pulsations in our
simulations is twofold. It is the external normally-distributed random
forces and the mutual interparticle interactions leading to
randomization as well. Speaking about turbulence, first of all it is
necessary to scale the fluctuations. Using the average viscosity $\nu = 5\,\text{mm}^2/\text{s}$
and the mean enstrophy density of $\Omega = 0.4\,\text{s}^{-2}$
obtained in simulations, we get an
energy dissipation rate per unit mass \cite{Frisch1995} of the order of $\epsilon = 2 \nu \Omega =
4\,\text{mm}^2/\text{s}^3$. This results in a Kolmogorov length scale of $L_K = \left(\nu^3 / \epsilon \right) ^{1/4} = 2.4\,\text{mm}$,
which is an intermediate scale between the size of the simulation box
($20 \times 45\,\text{mm}^2$) and the interparticle distance ($0.15\,$mm).

\begin{figure}
  \includegraphics[width=.9\columnwidth]{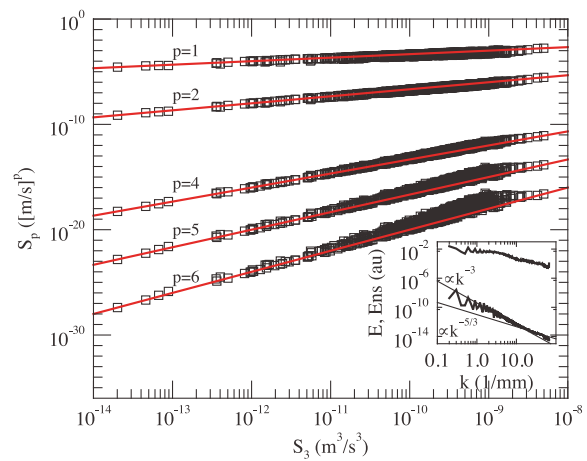} %,bb= 40 210 550 600
  \caption{\label{fig:Sp}(color online) Squares: First (p=1) through sixth (p=6)
    order longitudinal velocity structure functions $S_p$ as a function of the third
    order structure function $S_3$. The lines are those predicted by
    classical Kolmogorov theory $S_p \propto r^{\zeta_p}$, with
    $\zeta_p = p/3$. Inset: Energy (bottom) and Enstrophy density
    spectrum (top). The slopes of the lines overplotted on the energy
    spectrum are -3 and -5/3.}
\end{figure}

Traditionally, the scaling of vortices is studied by computing
structure functions. The longitudinal structure functions of order $p$, $S_p (r)$, are given by
\begin{equation}
  S_p(r) = \left< ( \Delta u_r )^p \right> = \left< [ u(x+r) - u(x)]^p \right>,
\end{equation}
where $u(x)$ is the velocity component parallel to the relative
displacement r. We use the absolute value of the velocity difference
in the calculations, as this is more numerically stable and does not
significantly change the results \cite{Lewis1999}. For fully
developed turbulence, the structure functions show a power law scaling
\cite{Frisch1995}
\begin{equation}
  S_p \propto r^{\zeta_p}.
\end{equation}
\citet{Kolmogorov1941} predicted the exponents to be given by $\zeta_p
= p/3$. Figure~\ref{fig:Sp} shows the horizontal velocity structure functions
measured in the region to the right of the void, where the particles
are flowing mainly in horizontal direction. As recommended by
\citet{Lewis1999}, we use extended self-similarity \cite{Benzi1993},
i.e., we plot $S_p$ vs. $S_3$ on a log-log plot and use the fact that
$\zeta_3 = 1$ according to Kolmogorov's theory. The slopes of the plots
are then given by $\zeta_p / \zeta_3 = \zeta_p$. Figure~\ref{fig:Sp}
shows that the slopes are very close to those predicted by Kolmogorov
(as the overplotted solid lines in the figure indicate).

The inset in Fig.~\ref{fig:Sp} shows the energy and enstrophy density
spectra. The energy spectrum is calculated from the sum of the squares of the Fourier
transformed velocity maps (see Fig.~\ref{fig:velocities}), the enstrophy
spectrum from the enstrophy map (see Fig.~\ref{fig:enstrophy}). The
two lines that are overplotted on the energy spectrum have the slopes
-3 and -5/3, as expected: The $\propto k^{-5/3}$ dependence is typical
inside the inertial range, and the
cross-over to $\propto k^{-3}$ is due to friction \cite{Frisch1995}.

{\it Summary---}To conclude, we numerically studied two-dimensional vortices in a
complex plasma. We showed how these vortices develop from a monolayer
around the void, and that a non-zero curl of the combined ion drag and electric
forces causes the vortex movement of the microparticles. We then
investigated the properties of fully developed system of vortices, for instance the flow
field and shear stress. In particular, we showed that turbulence is
present in the flow induced by the vortices and demonstrated that the
velocity structure functions scale very close to the predictions by
Kolmogorov theory. These results show that it should be possible to use common experimental situations
like vortices in complex plasmas to study turbulence on the level of
individual particles.

\begin{acknowledgments}
  We acknowledge support by a Marie Curie International Outgoing
  Fellowship within the 7th European Community Framework Programme,
  from the European Research Council under the European Union’s
  Seventh Framework Programme (FP7/2007- 2013)/ERC Grant Agreement
  No. 267499, and by the US Department of Energy, Office of Fusion
  Science Plasma Science Center.  The corresponding
experiments on the International Space Station were funded by DLR/BMWi
under the contract numbers FKZs 50 WM 0203 and 50 WM 1203.
\end{acknowledgments}

%\bibliography{/Users/schwabe/Documents/bas/bibliography.bib}

\begin{thebibliography}{41}
\bibitem{footnote1}{The enhancement of the borders of the cloud is an
artefact of the calculation (we only calculate the force for positions
with microparticles -- outside of the cloud, the total force is thus
zero in our calculation. The algorithm to calculate the curl acts as
an edge detection algorithm as the gradient is largest where the force
falls to zero.) }
\bibitem{footnote2}{As is easy to check, this also readily follows
    from a simplified rotation balance: $|\nabla \times {\bf v}| \sim
    (m \gamma_{\text{eff}})^{-1} |\nabla \times {\bf F}|$, where
    $\gamma_{\text{eff}} = \gamma_{\text{Ep}} + \nu k_{\text{eff}}^2$
    with $k_{\text{eff}} \sim 2 \pi / L_K$.}
\bibitem{footnote3}{Note that in experiments, more complicated situations exist as well, for
instance a reversal of the direction of rotation or a system with
several vortices, which we have not observed in simulations so far.}
\bibitem{footnote4}{Values with $\gamma^2 < 0.05\,\text{s}^{-2}$ were
omitted in the plot, as the viscosity cannot be calculated by the
above formula for $\gamma^2$ that are too small \cite{Nosenko2013}.}
\end{thebibliography}

\end{document}